\documentclass[
aip,
jcp,
numerical,
amsmath,amssymb,
preprint
]{revtex4-1}

\usepackage{graphicx}
\usepackage{dcolumn}
\usepackage{bm}
\usepackage{siunitx}
\usepackage{multirow}
\usepackage{hyphenat}

\begin{document}

\title[Space partitioning of exchange-correlation functionals with the projector augmented-wave method]{Space partitioning of exchange-correlation functionals with the projector augmented-wave method}

\author{H. Lev\"am\"aki}%
 \email{levamaki@kth.se}
\affiliation{Applied Materials Physics, Department of Materials Science and Engineering, Royal Institute of Technology, Stockholm SE-100 44, Sweden}%
\author{M. Kuisma}%
\affiliation{ 
Nanoscience Center, Department of Chemistry, University of Jyv\"askyl\"a, P.O. Box 35, FIN-40014, Jyv\"askyl\"a, Finland
}%

\author{K. Kokko}
\affiliation{Department of Physics and Astronomy, University of Turku, FI-20014 Turku, Finland
}%
\affiliation{Turku University Centre for Materials and Surfaces (MatSurf), FI-20014 Turku, Finland
}%

\date{\today}

\begin{abstract}
We implement a Becke fuzzy cells type space partitioning scheme for the purposes of exchange-correlation within the GPAW projector augmented-wave method based density functional theory code.
Space partitioning is needed in the situation where one needs to treat different parts of a combined system with different exchange-correlation functionals.
For example, bulk and surface regions of a system could be treated with functionals that are specifically designed to capture the distinct physics of those regions.
Here we use the space partitioning scheme to implement the quasi-nonuniform exchange-correlation scheme, which is a useful practical approach for calculating metallic alloys on the generalized gradient approximation level.
We also confirm the correctness of our implementation with a set of test calculations.
\end{abstract}


\maketitle

\section{Introduction}
Density functional theory (DFT) \cite{Hohenberg1964,Kohn1965} has become a standard technique for the computation of electronic properties of molecules and periodic systems.
As a theory DFT is exact, but in practical calculations one of the functionals, the exchange-correlation (XC) functional, has to be always approximated.
One of the great strengths of DFT is that even the simplest XC approximation, the local density approximation (LDA) \cite{Kohn1965,Vosko1980,Perdew1981,Perdew1992}, turned out to be remarkably useful, albeit mostly for physics related applications and to a lesser degree for chemistry.
The next family of XC approximations beyond the LDA are generalized gradient approximations (GGAs), which offer fairly ubiquitous improvements over LDA.
One of the most well-known GGA XC functionals is the Perdew-Burke-Ernzerhof (PBE) functional \cite{Perdew1996a}, which has earned itself the rank of a ``standard functional''.
Though even more sophisticated XC approximations, such as meta-GGAs \cite{Ghosh1986,Becke1989,Tao2003} and hybrid XC functionals \cite{Becke1993a,Becke1993}, are being developed and further improved, GGAs are still the most sensible choice for many applications due to their excellent computational speed versus accuracy ratio.

The potential accuracy of GGAs, however, has a certain ceiling, because the information that a GGA functional has about any given system is limited to the electron density $n$ and the gradient of the density $\nabla n$.
This deals with the computation of periodic and solid-state systems on GGA level, and in this realm there exists evidence that we might be approaching the accuracy limit.
For example, a recent paper by Tran \textit{et al.} \cite{Tran2016c} benchmarks various GGAs on their accuracy for a set of important solid-state properties. 
Equilibrium lattice constant is one the most fundamental properties in solid-state physics, and Fig. 4 of the work of Tran \textit{et al.} shows that modern solid-state GGAs (WC \cite{Wu2006}, PBEsol \cite{Perdew2008}, SOGGA \cite{Zhao2008}, and SG4 \cite{Constantin2016e}) are all clustering above a mean absolute relative error (MARE) of about $0.5\%$.
This suggests that the MARE of $\sim0.5\%$ represents an accuracy barrier that has proved difficult to breach on the GGA level.
The accuracy limit of GGA functionals has important consequences.
In the work of Tian \textit{et al.} \cite{Tian2016prl} it has been discussed why GGA functionals often yield poor formation energies for metallic binary alloys, and the reason is connected to the GGA lattice constant accuracy limit.
An accurate formation energy requires that the equation of states (and therefore lattice constants) of all alloy components are described accurately.
On GGA level this is often impossible. 

In this paper we utilize the ``fuzzy cells'' space partitioning concept of Becke \cite{Becke1988} to circumvent the accuracy limit of the GGA level by implementing a space partitioned GGA XC functional QNA \cite{Levamaki2012,Levamaki2014} within the state-of-the-art electronic structure code GPAW \cite{Mortensen2005,Enkovaara2010}.
Our starting point is the PBE-family (PBE, PBEsol), whose functionality is governed by two parameters noted as $\mu$ and $\beta$.
The parameter $\mu$ gives the strength of GGA corrections over LDA exchange and $\beta$ gives the strength of corrections over LDA correlation.
Thus, in general, accurate description of XC effects only with semi-local description of PBE-ansatz with energy functional $E_{\rm xc}[n({\bf r});\mu,\beta]=\int {\rm d} {\bf r}
n({\bf r})
\epsilon( n({\bf r}), |\nabla n({\bf r})|^2, \mu,\beta)$ is not possible, where
$\epsilon$ is the XC energy per particle.
A generic element within chemistry is obviously a single atom, and without any external fields, the external potential $v_{\rm ext}({\bf r})$ is solely a function of the atomic positions. Especially in case of solid alloys with metallic bonds we consider improving GGA by explicitly parametrising it in a volume around each atom species.
In this approach, called quasi-nonuniform approximation (QNA) \cite{Levamaki2012,Levamaki2014}, the XC functional is no longer a density functional theory in a strict sense, but becomes also a function of atomic positions and information of species
$E^{\rm QNA}_{\rm xc}[n;\{ ({\bf R}^a, \mu^a, \beta^a) \}]=\int {\rm d}{\bf r} n({\bf r}) \epsilon_{\rm QNA}( n({\bf r}), |\nabla n({\bf r})|^2, \{ ({\bf R}^a, \mu^a, \beta^a) \} )$, where ${\bf R}^a$ is the atomic positions and $\mu^a$ and $\beta^a$ are atom-specific parameters described later in the text. The approach has been previously implemented in exact muffin-tin orbitals (EMTO) method \cite{andersen1994lectures,Vitos2001,Vitos2001c,Vitos2007b} and good results has been obtained for various binary alloys \cite{Tian2016prl}. However, in the original implementation volumes with a strict Voronoi partition were used, rendering the local PBE-ansatz parameters discontinuous with respect to ${\bf r}$. Here we overcome the difficulty by employing the fuzzy cells space partitioning concept of Becke \cite{Becke1988}, which allows the computation of analytic QNA forces and stress tensor.
For efficient calculations, we implement the projector augmented wave method corrections \cite{Blochl1994,Kresse1999} to QNA within the projector augmented-wave method based DFT code GPAW. Atomic Simulation Environment (ASE) \cite{Bahn2002,HjorthLarsen2017} is used thorough the article for handling the atomic geometries and optimizations.

\section{Implementation} \label{sec:implementation}
The QNA scheme essentially generalizes the $\mu$ and $\beta$ parameters of PBE XC functional \cite{Perdew1996a} into space dependent $\mu({\bf r})$ and $\beta({\bf r})$ fields
\begin{align}
\mu({\bf r}) = \sum_a w_a({\bf r}) \mu^a, 
\label{eq:mu} \\
\beta({\bf r}) = \sum_a w_a({\bf r}) \beta^a,
\end{align}
where $\mu^a$ and $\beta^a$ are optimized parameters corresponding to a given element occupying atomic site $a$.
Consequently, the QNA XC energy can be written in the form
\begin{equation}
E^{\rm QNA}_{\rm XC}[n]
= \int 
n({\bf r})\varepsilon_{\rm XC}^{\rm PBE}[n({\bf r}), |\nabla n({\bf r})|^2, \mu({\bf r}), \beta({\bf r})]\,{\rm d} {\bf r},
\end{equation}
where 
\begin{equation}
\epsilon^{\rm PBE}_{\rm XC} = \varepsilon_{\rm X}^{\rm LDA} \left(
F_{\rm X}^{\rm PBE}[n({\bf r}), |\nabla n({\bf r})|^2, \mu({\bf r})] +
\left\{ \frac{\varepsilon_{\rm C}^{\rm LDA}}{\varepsilon_{\rm X}^{\rm LDA}} + \frac{H[n({\bf r}), |\nabla n({\bf r})|^2, \beta({\bf r})]}{\varepsilon_{\rm X}^{\rm LDA}}\right\} \right)
\end{equation}
is the PBE-type XC energy density  per particle.
The $\mu({\bf r})$ and $\beta({\bf r})$ fields should interpolate sharply between atoms and in practice this creates the need to divide space into Voronoi-type atomic site centered regions.
Space division can be accomplished by appropriate weight fields $w_a({\bf r})$. 
The value of $w_a({\bf r})$ should approach unity close to atomic site $a$, and decay smoothly to zero away from site $a$.
Additionally, it must always hold that $\sum_a w_a({\bf r}) = 1$. 
We define the weights as
\begin{equation} \label{eq:Becke_voronoi}
w_a({\bf r}) = \frac{ P_a({\bf r}) }{ \sum_{a'} P_{a'}({\bf r}) },
\end{equation}
which follows the fuzzy cells concept first developed by Becke \cite{Becke1988}.
In the fuzzy cells scheme $P_a({\bf r})$ are atomic site centered partial weights, which have the value one at the atomic site ${\bf R}^a$ and decay to zero when the distance $|{\bf r} - {\bf R}^a|$ becomes large.
$P_a({\bf r})$ could be defined many different ways, but here we will use
\begin{equation} \label{eq:Pa}
P_a({\bf r}) = f(|{\bf r} - {\bf R}^a|) = 
\exp{ \left[ - \left(\frac{|{\bf r} - {\bf R}^a|}{\lambda}\right)^{2\alpha} \right]},
\end{equation}
which is very similar to the expression developed in Ref. \onlinecite{Franchini2013}.
The parameter $\lambda$ controls the location of the transition from 1 to 0 and $\alpha$ controls the sharpness of the transition.
We have found that values $\lambda=1.2$ and $\alpha=2.0$ give partitioning that is very close to exact Voronoi cells, and also the most accurate formation energies. The calibration of formation energies has been done by calculating the formation energies of ordered Cu$_3$Au and CuAu$_3$ (L$1_2$), and CuAu (L$1_0$) and then comparing them to previous EMTO QNA results \cite{Tian2017CuAu}.

For periodic and solid-state calculation the expression of Eq. (\ref{eq:Pa}), and that of Ref. \onlinecite{Franchini2013}, for $P_a({\bf r})$ is particularly beneficial, because the computational load of Eq. (\ref{eq:Pa}) scales only linearly as a function of nuclei.
This is in contrast to the quadratic scaling of the original Becke form and others \cite{Salvador2013}, which are often used in chemistry.
Chemistry calculations routinely employ computationally heavy hybrid XC functional, which means a quadratic scaling $P_a({\bf r})$ is responsible for only a fraction of the total computational load.
However, in solid state physics fast semilocal LDA and GGA XC functionals are popular, which can easily cause a quadratic scaling $P_a({\bf r})$ to become a computational bottleneck.

Performing geometric relaxations using the QNA scheme requires the computation of forces and the stress tensor. For all-electron case, the XC potential can be  evaluated in the usual fashion as
\begin{equation}
v_{\rm xc}({\bf r}) = \frac{\delta 
E_{\rm XC}^{\rm PBE}[n({\bf r}), \mu({\bf r}), \beta({\bf r})]}{\delta n({\bf r})},
\label{vxc}
\end{equation}
where the dependence on  $\mu({\bf r})$ and $\beta({\bf r})$ is purely parametric as they are not explicit functions of density but of nuclear coordinates. This equation is now in useful form, as it allows simple analytical gradients.

We now consider the projector augmented wave method implementation and begin with a brief introduction of relevant concepts. The general idea in PAW-methods is that the Kohn--Sham equations are solved for smooth wave functions ($\tilde \psi_n({\bf r})$), but retaining one-to-one mapping with the all-electron wave functions ($\psi({\bf r})$). There exists a linear PAW transformation operator which defines a mapping $\psi({\bf r}) = \mathcal{T} \tilde \psi_n({\bf r})$ and the Kohn--Sham equations are derived to be $
    \mathcal{T}^\dagger H \mathcal{T}\tilde \psi_n({\bf r}) =\mathcal{T}^\dagger \mathcal{T} \tilde \psi_n({\bf r})
$. Furthermore, several pseudo quantities are defined, such as the pseudo charge density
$\tilde \rho({\bf r})$ and 
the pseudo density ($\tilde n_\sigma({\bf r})$) which is written as (we consider here a spin-paired system for simplicity and drop the spin index)

\begin{equation}
    \tilde n({\bf r}) = \tilde n_c({\bf r}) + \sum_i f_i |\tilde \psi_n({\bf r})|^2.
    \label{pseudodens}
\end{equation}

In GPAW code, in all of the grid, LCAO and plane wave modes, this density is defined in Cartesian real space grid with grid-spacing typically between 0.07-0.15 \AA. Furthermore, around each nucleus $a$, one can define the pseudo ($\tilde n^a({\bf r})$) and all-electron densities ($n^a({\bf r})$) which are defined in logarithmic radial grid where the angular part is expanded using 50 Lebedev points. For $r>r_c$, where $r_c$ is the PAW cutoff of an atomic sphere, it holds that $\tilde n^a({\bf r}) = n^a({\bf r})$ with their derivatives also matched. The crux of PAW implementation of QNA, is to define analogous quantities $\tilde \mu({\bf r})$, $\tilde \mu^a({\bf r})$ and $\mu^a({\bf r})$ and $\tilde \beta({\bf r})$, $\tilde \beta^a({\bf r})$, and $\beta^a({\bf r})$ respectively. Although the end result is simpler than this, we utilize these quantities when deriving to quantify the approximations made. For now, the three versions per $\mu$ and $\beta$ fulfill the same rules as density quantities $\tilde n$, $\tilde n^a$, and $n^a$.

The energy gradients in PAW formalism can be in general written in form 
\begin{equation}
    {\bf F}^a = -\frac{{\rm d} E}{{\rm d}{\bf R}^a} = -\frac{\partial E}{\partial {\bf R}^a}
    - \sum_n \left(
    \int {\rm d}{\bf r}
    \frac{\partial E}{\partial \tilde \psi_n({\bf r})}
    \frac{{\rm d} \tilde \psi_n}{{\rm d}{\bf R}^a}
    + {\rm h.c.} \right) =
    -\frac{\partial E}{\partial {\bf R}^a}
    + \sum_n f_n \epsilon_n \langle \tilde \psi_n| \frac{{\rm d} \hat S}{{\rm d} {\bf R}^a} |\tilde \psi_n\rangle,
\end{equation}
where ``h.c.'' denotes the Hermitian conjugate.
In case of the QNA XC functional, it does not have explicit wave function dependence and hence we only need to consider the partial derivative $-\frac{{\rm d} E}{{\rm d}{\bf R}^a}$.
For local and semi-local functionals, the XC energy in PAW formalism can be written as
\begin{equation}
    E_{\rm XC} =
    E_{\rm XC}[\tilde n({\bf r}), \tilde \mu({\bf r}), \tilde \beta({\bf r})]  + \sum_a
    E^a_{\rm XC}[n^a({\bf r}), \mu^a({\bf r}), \beta^a({\bf r})]
    -E^a_{\rm XC}[\tilde n^a({\bf r}), \tilde \mu^a({\bf r}), \tilde \beta^a({\bf r})]
    \label{Excpaw}
\end{equation}
where the term
\begin{equation}
\Delta E^a = E_{\rm XC}[n^a[D^a_{ii'}], \mu^a({\bf r}), \beta^a({\bf r})]
    -E_{\rm XC}[\tilde n^a[D^a_{ii'}],\tilde \mu^a({\bf r}), \tilde \beta^a({\bf r})]
\end{equation}
is typically called the PAW-correction and it introduces the atomic density matrix as defined in Ref. \onlinecite{PhysRevB.80.195112}.
By taking the partial derivative with respect to nuclear position of Eq.~(\ref{Excpaw}), 
we arrive at Eq. (\ref{eq:dE_QNA_dR}), which is presented in Appendix A.
The first term in the right hand side of Eq. (\ref{eq:dE_QNA_dR}) is already handled by GPAW, and it is solved by noting that in Eq.~(\ref{pseudodens}), only the pseudo core density $\tilde n_c({\bf r})$ depends on atomic positions i.e.
\begin{equation}
   -
\int {\rm d}{\bf r}
\frac{\delta
E_{\rm XC}[\tilde n] }{\delta \tilde n({\bf r})}
\frac{\partial \tilde n({\bf r})}{\partial {\bf R}^a}
=
   -
\int {\rm d}{\bf r}
\frac{\delta
E_{\rm XC}[\tilde n] }{\delta \tilde n({\bf r})}
\frac{\partial \tilde n_c({\bf r})}{\partial {\bf R}^a}.
\label{term2}
\end{equation}
Furthermore, the density functional derivatives of the form $\delta E/\delta n$ in Eq.~(\ref{term2}) or via the atomic wise quantities in Eq.~(\ref{eq:dE_QNA_dR}) are readily evaluated in GPAW via the typical Euler-Lagrange derivation
\begin{align}
    v_{\rm XC}({\bf r}) &= 
    \frac{\delta E_{\rm XC}[n({\bf r})] }{\delta n({\bf r})} = 
    \frac{\partial \left( n({\bf r}) \epsilon_{\rm XC}[n({\bf r}), \nabla n({\bf r})] \right)}{\partial n({\bf r})}
    - \nabla \cdot \left(
    \frac{\partial \left( n({\bf r}) \epsilon_{\rm XC}[n({\bf r}),\nabla n({\bf r})]\right)}{\partial \nabla n({\bf r})}
    \right) \nonumber \\
    &= \frac{\partial \left(n({\bf r}) \epsilon_{\rm XC}[n({\bf r}), \nabla n({\bf r})]\right)}{\partial n({\bf r})}
    - \nabla \cdot \left(
    \frac{\partial \left(n({\bf r}) \epsilon_{\rm XC}[n({\bf r}), \sigma({\bf r})]\right)}{\partial \sigma({\bf r})}
     2 \nabla n({\bf r}) \right),
\end{align}
where $\sigma({\bf r})=|\nabla n({\bf r})|^2$.

Thus, regarding the new implementation of QNA forces, we are left with only the partial derivatives with respect to various $\mu$ and $\beta$ parameters.
Due to the atomic site centered $\mu({\bf r})$ and $\beta({\bf r})$ fields $E_{\rm XC}^{\rm QNA}$ has an additional dependency on the positions of the nuclei, which created the extra derivative chain rules in Eq. (\ref{eq:dE_QNA_dR}).
However, we have not yet defined the $\tilde \mu$, $\tilde \mu^a$ and $\mu^a$ (in the following, $\beta$ is defined analogously). To this end, we make a typical approximation, where a quantity almost constant within an augmentation sphere is assumed to be constant. In other words, we set $\tilde \mu^a({\bf r}) = \mu^a({\bf r}) = \mu^a$. Outside the augmentation sphere, the fact that $\tilde \mu^a({\bf r})$ deviates from $\tilde \mu^a$ does not matter, since the correction vanishes since $n^a({\bf r})=\tilde n^a({\bf r})$ there also. Inside the augmentation sphere, where $n^a({\bf r})$ and $n^a({\bf r})$ deviate, the region is so close to atom $a$ that $\mu^a$ term dominates in Eq.~(\ref{eq:mu}).

With these approximations, the QNA XC becomes

\begin{equation}
    E_{\rm XC} =
    E_{\rm XC}[\tilde n({\bf r}), \tilde \mu({\bf r}), \tilde \beta({\bf r})]  + \sum_a
    E^a_{\rm XC}[n^a({\bf r}), \mu^a, \beta^a]
    -E^a_{\rm XC}[\tilde n^a({\bf r}), \tilde \mu^a, \tilde \beta^a],
    \label{Excpaw2}
\end{equation}
and the forces simplify to
\begin{align} 
{\bf F}_{\rm XC}^a = 
&-
\int {\rm d}{\bf r}
\left(
\frac{\delta
E_{\rm XC}[\tilde n({\bf r}), \tilde \mu({\bf r}), \tilde \beta({\bf r})] }{\delta \tilde n({\bf r})}
\frac{\partial \tilde n({\bf r})}{\partial {\bf R}^a}
+\frac{\delta
E_{\rm XC}[\tilde n({\bf r}), \tilde \mu({\bf r}), \tilde \beta({\bf r})]}{\delta \tilde \mu({\bf r})}
\frac{\partial \tilde \mu({\bf r})}{\partial {\bf R}^a}\right. \nonumber \\
&+ \left.\frac{\delta
E_{\rm XC}[\tilde n({\bf r}), \tilde \mu({\bf r}), \tilde \beta({\bf r})]}{\delta \tilde \beta({\bf r})}
\frac{\partial \tilde \beta({\bf r})}{\partial {\bf R}^a} \right) \nonumber \\
 &-
\sum_a \sum_{ii'}
\int {\rm d}{\bf r}
\left(
\frac{\delta E^a_{\rm XC}[n^a({\bf r}), \mu^a, \beta^a]}{\delta n^a({\bf r})}
\frac{\partial n^a({\bf r})}{\partial D^a_{ii'}} +
\frac{\delta E^a_{\rm XC}[\tilde n^a({\bf r}), \mu^a, \beta^a]}{\delta \tilde n^a({\bf r})}
\frac{\partial n^a({\bf r})}{\partial D^a_{ii'}}
\right)
\frac{\partial D^a_{ii'}}{\partial {\bf R}^a}.
\label{eq:dE_QNA_dR2}
\end{align}
At this point we can readily evaluate the remaining partial derivatives
\begin{align}
&
\frac{\delta
E_{\rm XC}^{\rm PBE}}{\delta \tilde \mu({\bf r})}
\frac{\partial \tilde \mu({\bf r})}{\partial {\bf R}^a}
= 
\int 
\frac{\partial\left\{
\tilde n({\bf r})\varepsilon_{\rm XC}^{\rm PBE}[\tilde n({\bf r}), |\nabla \tilde n({\bf r})|^2, \tilde \mu({\bf r}), \tilde \beta({\bf r})]\right\}
}{\partial \tilde \mu({\bf r})}
\sum_{a'} \frac{\partial w_{a'}({\bf r})}{\partial {\bf R}^a} \mu^{a'} \,{\rm d} {\bf r}, \label{eq:dEXC_dmu} \\
&
\frac{\delta
E_{\rm XC}^{\rm PBE}}{\delta \tilde \beta({\bf r})}
\frac{\partial \tilde \beta({\bf r})}{\partial {\bf R}^a}
=
\int
\frac{\partial\left\{
\tilde n({\bf r})\varepsilon_{\rm XC}^{\rm PBE}[\tilde n({\bf r}), |\nabla \tilde n({\bf r})|^2, \tilde \mu({\bf r}), \tilde \beta({\bf r})]\right\}
}{\partial \tilde \beta({\bf r})}
\sum_{a'} \frac{\partial w_{a'}({\bf r})}{\partial {\bf R}^a} \beta^{a'} \,{\rm d} {\bf r}. \label{eq:dEXC_dbeta}
\end{align}
Inside Eqs. (\ref{eq:dEXC_dmu}) and (\ref{eq:dEXC_dbeta}) we have the further partial derivatives
\[
\frac{\partial\left\{
n\varepsilon_{\rm XC}^{\rm PBE}\right\}
}{\partial \tilde \mu({\bf r})},\quad
\frac{\partial\left\{
n\varepsilon_{\rm XC}^{\rm PBE}\right\}
}{\partial \tilde \beta({\bf r})}
\]
and they have been written out in Appendix A.
In order to get the $\delta w_{a'}({\bf r})/\delta {\bf R}^a$ derivatives in Eqs. (\ref{eq:dEXC_dmu}) and (\ref{eq:dEXC_dbeta}) we notice that
\begin{equation}
P_{a'}({\bf r}) = f \left(\sqrt{ 
    (r_x - R^{a'}_x)^2 +
    (r_y - R^{a'}_y)^2 +
    (r_z - R^{a'}_z)^2} \right),
\end{equation}
which gives, for example,
\begin{equation}
\frac{\partial P_{a'}({\bf r})}{\partial R^a_x} =
- \delta_{aa'} f' \left(\sqrt{\cdots} \right)
\frac{(r_x - R^{a}_x)}{| {\bf r} - {\bf R}^{a} |} =
- \delta_{aa'}\frac{\partial P_{a}({\bf r})}{\partial r_x}.
\end{equation}
The whole gradient with respect to ${\bf R}^a$ ($\nabla_{ {\bf R}^a }$) is therefore easily obtained from the gradient of ${\bf r}$ using the $\nabla$ operator:
\begin{equation} \label{eq:grad_transform}
\frac{\partial P_{a'}({\bf r})}{\partial {\bf R}^a}
= \delta_{aa'}\nabla_{ {\bf R}^a } P_{a}({\bf r})
= - \delta_{aa'}\frac{\partial P_{a}({\bf r})}{\partial {\bf r}}
= - \delta_{aa'}\nabla P_{a}({\bf r}).
\end{equation}
With the help of Eq. (\ref{eq:grad_transform}) we obtain the following expression for the $\delta w_{a'}({\bf r})$ $/$ $\delta {\bf R}^a$ derivatives:
\begin{align}
\frac{\partial w_{a'}({\bf r})}{\partial {\bf R}^a}
&= 
\frac{\partial}{
    \partial {\bf R}^a}
\frac{ P_{a'}({\bf r}) }{ \sum_{a''} P_{a''}({\bf r}) } \nonumber \\
&=
\frac{\frac{\partial P_{a'}({\bf r})}{
        \partial {\bf R}^a}
    \sum_{a''} P_{a''}({\bf r})
    - P_{a'}({\bf r}) 
    \sum_{a''}\frac{\partial P_{a''}({\bf r})}{\partial {\bf R}^a} 
}{ \left[\sum_{a''} P_{a''}({\bf r})\right]^2 } \nonumber \\
&=
\frac{
    - \delta_{aa'} \nabla P_a({\bf r})
    \sum_{a''} P_{a''}({\bf r})
    + P_{a'}({\bf r})
    \nabla P_{a}({\bf r})
}{ \left[\sum_{a''} P_{a''}({\bf r})\right]^2 }.
\end{align}

The stress tensor is needed in order to relax the unit cell.
Analogously to the case of forces, additional terms will manifest in the stress tensor formula, because the $\mu({\bf r})$ and $\beta({\bf r})$ fields change as a function of strain.
The derivations below make use of Ref. \onlinecite{Knuth2015} which explores the computation of various stress tensor contributions in detail.
Stress tensor $\sigma$ is defined as a first order change under a strain $\epsilon$ as
\begin{equation}
\sigma_{\alpha\beta} = \frac{1}{V}\frac{\partial E_{\rm tot}}{\partial \epsilon_{\alpha\beta}}.
\end{equation}
Since XC energy is an integral in real space, it can be shown that the XC contribution to the total stress tensor can be written as
\begin{equation} \label{eq:sigma_XC}
V\sigma^{\rm XC}_{\alpha\beta} = \frac{\partial E_{\rm XC}}{\partial \epsilon_{\alpha\beta}}
= \delta_{\alpha\beta}E_{\rm XC} + \int_V \frac{\partial [n({\bf r})\varepsilon_{\rm XC}({\bf r})]}{\partial \epsilon_{\alpha\beta}}\,{\rm d} {\bf r}.
\end{equation}
For simplicity, here we drop the pseudo ($\sim$) notation and just use generic $\mu({\bf r})$ and $\beta({\bf r})$.
Then the application of the chain rule to Eq. (\ref{eq:sigma_XC}) gives
\begin{widetext}
\begin{align}
&\int_V \frac{\partial \{n({\bf r})\varepsilon_{\rm XC}^{\rm PBE}[n({\bf r}), |\nabla n({\bf r})|^2, \mu({\bf r}), \beta({\bf r})]\}}{\partial \epsilon_{\alpha\beta}}\,{\rm d} {\bf r} \nonumber \\
=&\int_V \frac{\partial \{n({\bf r})\varepsilon_{\rm XC}^{\rm PBE}[n({\bf r}), |\nabla n({\bf r})|^2, \mu({\bf r}), \beta({\bf r})]\}}{\partial n({\bf r})}\frac{\partial n({\bf r})}{\partial \epsilon_{\alpha\beta}}\,{\rm d} {\bf r} \label{eq:QNA_chain_rule_2} \\
&+\int_V \frac{\partial \{n({\bf r})\varepsilon_{\rm XC}^{\rm PBE}[n({\bf r}), |\nabla n({\bf r})|^2, \mu({\bf r}), \beta({\bf r})]\}}{\partial \nabla n({\bf r})} \cdot \frac{\partial \nabla n({\bf r})}{\partial \epsilon_{\alpha\beta}}\,{\rm d} {\bf r} \label{eq:QNA_chain_rule_3} \\
&+\int_V \frac{\partial \{n({\bf r})\varepsilon_{\rm XC}^{\rm PBE}[n({\bf r}), |\nabla n({\bf r})|^2, \mu({\bf r}), \beta({\bf r})]\}}{\partial \mu({\bf r})}\frac{\partial \mu({\bf r})}{\partial \epsilon_{\alpha\beta}}\,{\rm d} {\bf r} \label{eq:QNA_chain_rule_4} \\
&+\int_V \frac{\partial \{n({\bf r})\varepsilon_{\rm XC}^{\rm PBE}[n({\bf r}), |\nabla n({\bf r})|^2, \mu({\bf r}), \beta({\bf r})]\}}{\partial \beta({\bf r})}\frac{\partial \beta({\bf r})}{\partial \epsilon_{\alpha\beta}}\,{\rm d} {\bf r}. \label{eq:QNA_chain_rule_5}
\end{align}
\end{widetext}

Below we expand each term one at a time (function arguments are dropped for simplicity). Equation (\ref{eq:QNA_chain_rule_2}) is the LDA\hyp{}level term and it can be written as
\begin{equation} \label{eq:QNA_LDA_term}
\int_V \frac{\partial \{n\varepsilon_{\rm XC}^{\rm PBE}\}}{\partial n}\frac{\partial n}{\partial \epsilon_{\alpha\beta}}\,{\rm d} {\bf r} = 
\int_V \frac{\partial n}{\partial \epsilon_{\alpha\beta}} \left[ \varepsilon_{\rm XC}^{\rm PBE} + n\frac{\partial \varepsilon_{\rm XC}^{\rm PBE}}{\partial n} \right]\,{\rm d} {\bf r}.
\end{equation}
Equation (\ref{eq:QNA_chain_rule_3}) is the GGA\hyp{}level gradient term and it can be written as
\begin{align} 
&\int_V \frac{\partial \{n\varepsilon_{\rm XC}^{\rm PBE}\}}{\partial \nabla n} \cdot \frac{\partial \nabla n}{\partial \epsilon_{\alpha\beta}}\,{\rm d} {\bf r} = 
\int_V n \frac{\partial \varepsilon_{\rm XC}^{\rm PBE}}{\partial \nabla n} \cdot \frac{\partial \nabla n}{\partial \epsilon_{\alpha\beta}}\,{\rm d} {\bf r} \nonumber \\
=&\int_V n \frac{\partial \varepsilon_{\rm XC}^{\rm PBE}}{\partial |\nabla n|^2} \left( \frac{\partial |\nabla n|^2}{\partial \nabla n} \right) \cdot \left( \frac{\partial \nabla n}{\partial \epsilon_{\alpha\beta}} \right)\,{\rm d} {\bf r} \nonumber \\
=&2\int_V n \frac{\partial \varepsilon_{\rm XC}^{\rm PBE}}{\partial |\nabla n|^2} \left( \nabla n \right) \cdot \left( \frac{\partial \nabla n}{\partial \epsilon_{\alpha\beta}} \right)\,{\rm d} {\bf r},
\end{align}
where we have used the fact that $\partial |\nabla n|^2 / \partial \nabla n = 2\nabla n$.
These LDA and GGA terms are already handled by GPAW.
Equation (\ref{eq:QNA_chain_rule_4}) arises from the fact that the $\mu({\bf r})$ field changes as a function of strain and it can be written as
\begin{align}
&\int_V \frac{\partial \{n\varepsilon_{\rm XC}^{\rm PBE}\}}{\partial \mu}\frac{\partial \mu}{\partial \epsilon_{\alpha\beta}}\,{\rm d} {\bf r} = 
\int_V n \frac{\partial \varepsilon_{\rm XC}^{\rm PBE}}{\partial \mu}
\sum_{a} \frac{\partial w_{a}}{\partial \epsilon_{\alpha\beta}} \mu^{a} \,{\rm d} {\bf r} \nonumber \\
=&\int_V n \varepsilon_{\rm X}^{\rm LDA} \frac{\partial F_{\rm X}^{\rm PBE}}{\partial \mu}
\sum_{a} \frac{\partial w_{a}}{\partial \epsilon_{\alpha\beta}} \mu^{a} \,{\rm d} {\bf r}.
\end{align}
Equation (\ref{eq:QNA_chain_rule_5}) is the $\beta({\bf r})$ field change and it can be written as
\begin{align}
&\int_V \frac{\partial \{n\varepsilon_{\rm XC}^{\rm PBE}\}}{\partial \beta}\frac{\partial \beta}{\partial \epsilon_{\alpha\beta}}\,{\rm d} {\bf r} = 
\int_V n \frac{\partial \varepsilon_{\rm XC}^{\rm PBE}}{\partial \beta}
\sum_{a} \frac{\partial w_{a}}{\partial \epsilon_{\alpha\beta}} \beta^{a} \,{\rm d} {\bf r} \nonumber \\
=&\int_V n \frac{\partial H}{\partial \beta}
\sum_{a} \frac{\partial w_{a}}{\partial \epsilon_{\alpha\beta}} \beta^{a} \,{\rm d} {\bf r}.
\end{align}
$\partial F_{\rm X}^{\rm PBE} / \partial \mu$ and $\partial H / \partial \beta$ terms were already derived in the Forces section.
The $\partial w_{a} / \epsilon_{\alpha\beta}$ derivative can be written as
\begin{equation}
\frac{\partial w_{a'}}{\partial \epsilon_{\alpha\beta}} =
\frac{\partial}{
\partial \epsilon_{\alpha\beta}}
\frac{ P_{a} }{ \sum_{a'} P_{a'}}
=
\frac{\frac{\partial P_{a}}{
\partial \epsilon_{\alpha\beta}}
\sum_{a'} P_{a'} - P_{a} \sum_{a'} \frac{\partial P_{a'}}{\partial \epsilon_{\alpha\beta}} 
}{ \left[\sum_{a'} P_{a'}\right]^2 }.
\end{equation}
To get the $\partial P_{a} / \epsilon_{\alpha\beta}$ derivative we use Eq. (15) of Ref. \onlinecite{Knuth2015}:
\begin{align}
\frac{\partial P_a}{\partial \epsilon_{\alpha\beta}} =& \frac{\partial f(|{\bf r} - {\bf R}^a|)}{\partial \epsilon_{\alpha\beta}} = \frac{\partial f(|{\bf r} - {\bf R}^a|)}{\partial r_\alpha}(r_\beta - R^a_\beta) \\
=&f'(|{\bf r} - {\bf R}^a|)\frac{r_\alpha - R^a_\alpha}{|{\bf r} - {\bf R}^a|}(r_\beta - R^a_\beta) =
\frac{\partial P_a}{\partial r_\alpha}(r_\beta - R^a_\beta).
\end{align}

The above equations can be used to implement the needed stress tensor corrections in GPAW, or any other ``stress tensor compatible'' DFT code for that matter, but it has been shown that terms like those of Eq. (\ref{eq:QNA_chain_rule_4}) and Eq. (\ref{eq:QNA_chain_rule_5}), i.e. ones that are a consequence of the fact that the space has been partitioned, seem to be so small that they fall below the general numerical accuracy of DFT codes \cite{Knuth2015}.

\section{Test calculations}
\begin{figure}[h!]
\centering
\includegraphics[width=0.9\textwidth]{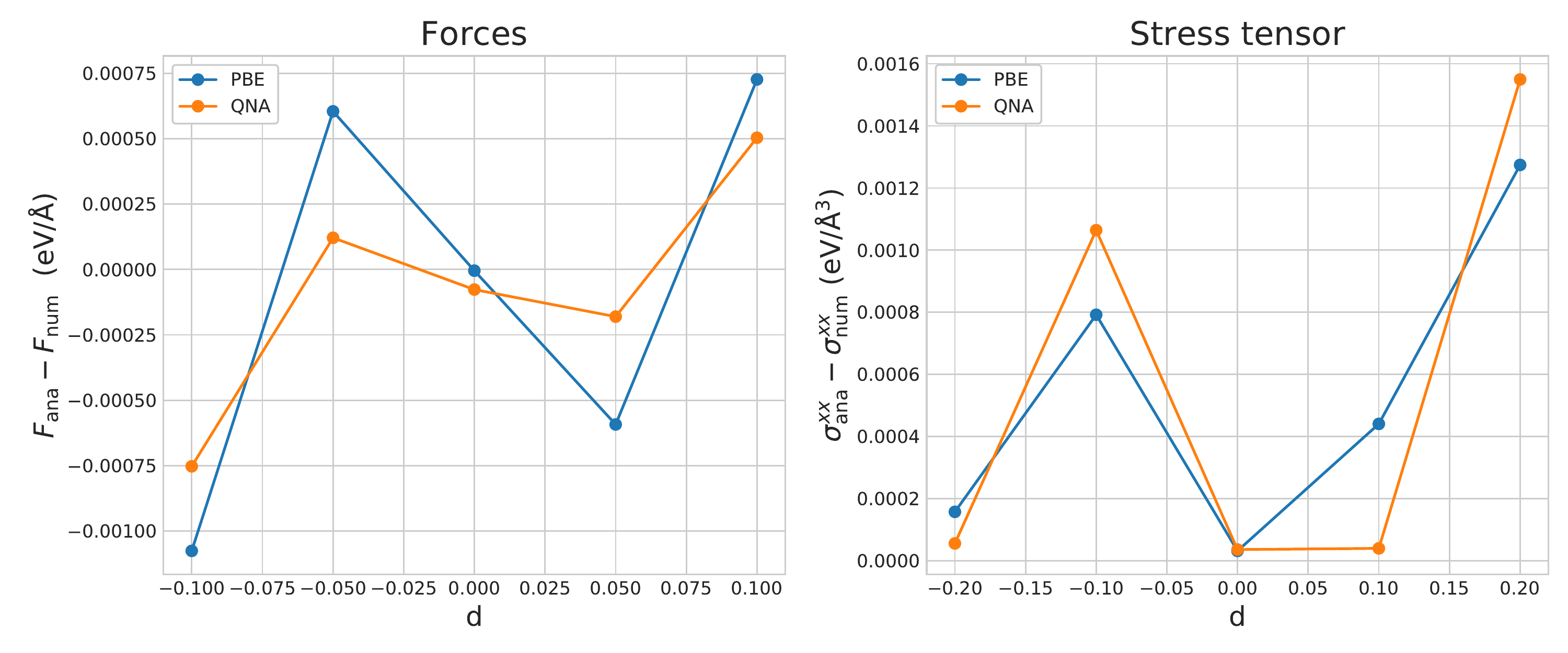}
\caption{The difference between analytically and numerically calculated forces and stress tensor for L$1_2$ Cu$_3$Au. For forces $d$ means that the Au atom at (0,0,0) lattice coordinate has been shifted by $d$ along the $x$-axis.
For the $xx$ component of the stress tensor $d$ means that the unit cell vector $a_1$ has been increased by $d$.}
\label{fig:force_stress_diff}
\end{figure}

The correctness of the analytical QNA forces and stress tensor can be straightforwardly checked by comparing them against numerically calculated finite-difference forces and stress tensor.
For example, the numerical force of atom $a$ in $x$-direction can be computed by displacing the atom by $\pm d$ along the $x$-direction and then calculating the finite difference $[E(+d) - E(-d)]/2d$.
As another example, the numerical $\sigma^{xx}$ component of the stress tensor is similarly computed by stretching the unit cell vector $a_1$ by $\pm d$ and then taking $[E(+d) - E(-d)]/2dV$, where $V$ is the unit cell volume.
Figure \ref{fig:force_stress_diff} shows the differences between analytical and numerical forces and $\sigma^{xx}$ stress tensor component for L$1_2$ Cu$_3$Au.
In the figure $d$ gives the displacement of the Au atom from its $(0,0,0)$ ideal lattice position along the $x$-axis.
For the $\sigma^{xx}$ stress tensor component $d$ tells by how much the optimized lattice vector $a_1$ has been increased/decreased along the $x$-axis.
We see that the differences between analytical and numerical QNA forces and stress tensor are very similar to those of PBE calculated with an unadulterated version of the GPAW code.
This confirms that the equations derived above for the analytical QNA forces and stress tensor work as expected.
\begin{table}[h!]
\centering
\caption{Formation energies of Cu-Au binary alloys. The VASP, EMTO, and experimental results are from literature and GPAW results are calculated using the implementation of this paper.}
\begin{tabular}{l l l l l} 
 \hline \hline
& Cu$_3$Au (L$1_2$) & CuAu (L$1_0$) & CuAu$_2$ ($\beta_2$) & CuAu$_3$ (L$1_2$) \\ \hline
PBE (VASP) \cite{Zhang2014prl} & $-44$ & $-56$ & $-44$ & $-25$ \\
PBE (EMTO) \cite{Tian2016prl} & $-45$ & $-57$ & \dots & $-24$ \\
PBE (GPAW) & $-40$ & $-52$ & $-41$ & $-21$ \\ \hline
QNA (EMTO) \cite{Tian2016prl} & $-70$ & $-87$ & \dots & $-41$ \\
QNA (GPAW) & $-71$ & $-85$ & $-61$ & $-42$ \\ \hline
Exp. \cite{Zhang2014prl} & $-74$ & $-93$ & \dots & $-39$ \\
 \hline \hline
\end{tabular}
\label{table:CuAu_results}
\end{table}

As a first practical test of our implementation we calculate the formation energies of ordered Cu-Au binary alloys.
The Cu-Au system is a famous prototype in alloy theory and it has been shown that GGA-level functionals struggle to predict the formation energies of Cu-Au binary alloys with acceptable accuracy \cite{Zhang2014prl,Tian2016prl}.
Table \ref{table:CuAu_results} shows the formation energies of Cu-Au binary alloys calculated with PBE and QNA using the present GPAW implementation.
We see that the GPAW results for QNA are in good agreement with previously published EMTO results.
The present GPAW implementation differs from the EMTO implementation in the way that in EMTO the space is by construction divided into Voronoi-cells that surround the muffin-tin spheres, and therefore does not need the fuzzy cells formalism.
Nevertheless, the results between the two codes agree, which indicates that QNA results are not sensitive to the underlying implementation and that the stress tensor can be succesfully used with QNA to optimize the unit cell geometry.

\begin{figure}[t]
\centering
\includegraphics[width=0.7\textwidth]{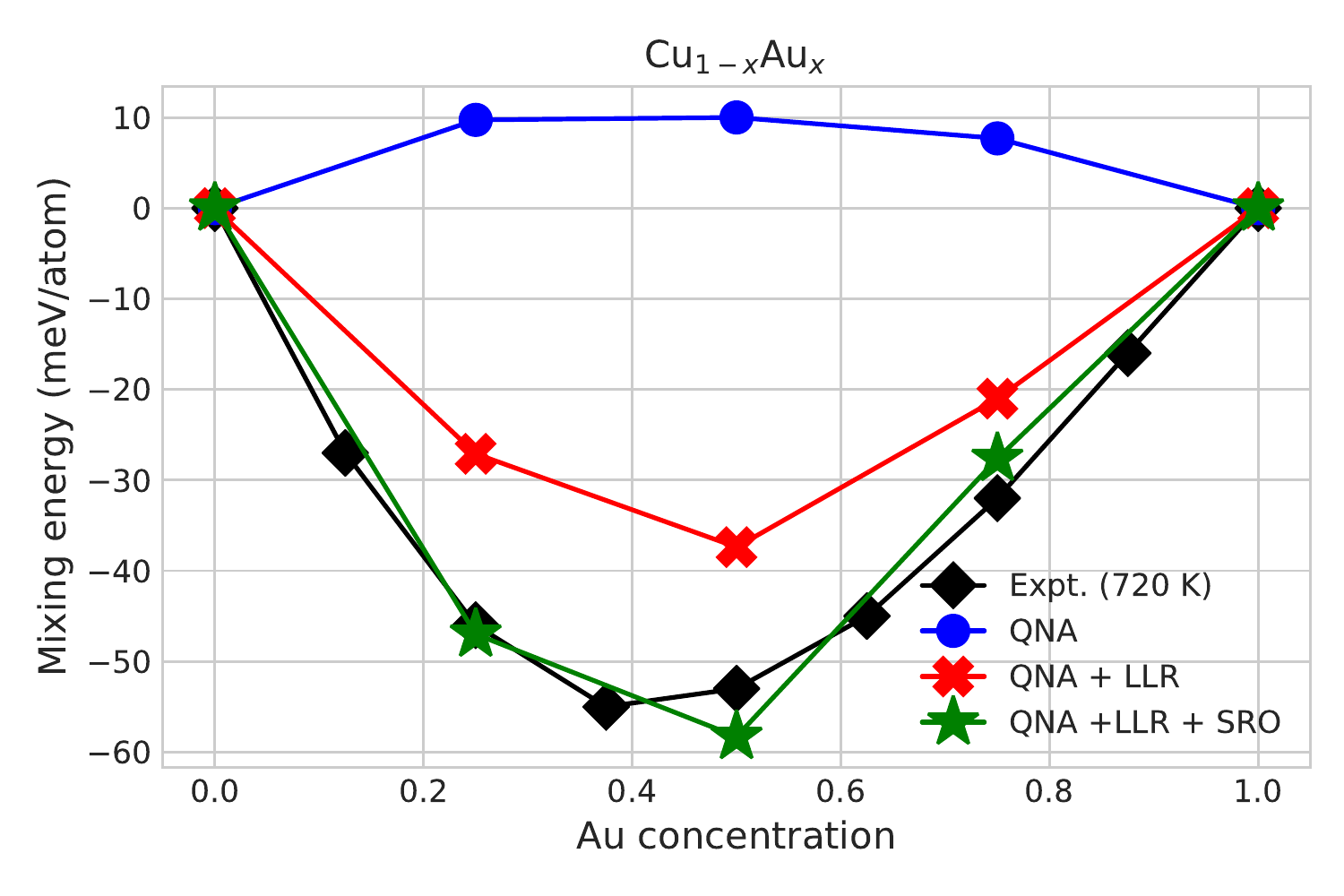}
\caption{Mixing energies of random Cu$_{1-x}$Au$_x$ alloys without LLRs (QNA), with LLRs (QNA+LLR), and with LLRs and a short-range order correction (QNA+LLR+SRO). Experimental values are from Ref. \onlinecite{Orr1960a}.}
\label{fig:CuAu_SQS}
\end{figure}
Next, in order to test the implementation of QNA forces, we calculate mixing energies of random Cu$_{0.75}$Au$_{0.25}$, Cu$_{0.5}$Au$_{0.5}$, and Cu$_{0.25}$Au$_{0.75}$ alloys using 32-atom special quasirandom structures (SQS) \cite{Zunger1990,Wei1990a} generated with the ATAT package \cite{avdw_atat,avdw_atat2,avdw_mcsqs}.
Previous studies have found that local lattice relaxations (LLRs) are very important in Cu-Au alloys due to the large atomic size mismatch between Cu and Au atoms \cite{Tian2017CuAu}.
For random Cu-Au alloys it is therefore important to be able to relax the atomic coordinates using forces.
Figure \ref{fig:CuAu_SQS} shows mixing energies of Cu$_{1-x}$Au$_x$ alloys as a function of $x$.
It can be seen that the mixing energies without LLRs (label QNA in figure) are positive and therefore qualitatively wrong.
Mixing energies with LLRs (QNA+LLR) are much improved, and by adding a short-range order estimate at the experimental temperature from Ref. \onlinecite{Tian2017CuAu} we arrive at values (QNA+LLR+SRO) that are very close to the experimental values of Ref. \onlinecite{Orr1960a}.
The good agreement of the ``QNA+LLR+SRO'' with experiments confirms that the QNA forces and the stress tensor are calculated correctly and with good accuracy.

\begin{table}[t]
\centering
\caption{Formation energies $\Delta E$ (in meV/atom) and magnetic moments $M_{\rm Fe}$/$M_{\rm Fe}$ (in $\mu_B$) of ferromagnetic and paramagnetic Fe$_3$Pt alloy.}
\begin{tabular}{c c c c c c} 
 \hline \hline
Fe$_3$Pt & Magn. & XC & $\Delta E$ & $M_{\rm Fe}$ & $M_{\rm Pt}$\\ \hline
& \multirow{4}{*}{FM} & LDA & $-22$ & $1.66$ & $0.10$ \\ 
&  & PBE & $-80$ & $2.77$ & $0.34$ \\ 
&  & PBEsol & $-50$ & $2.62$ & $0.30$ \\ 
&  & QNA & $-110$ & $2.67$ & $0.32$ \\ \hline
& \multirow{4}{*}{PARA} & LDA & $-21$ & $0.0$ & $0.0$ \\ 
&  & PBE & $24$ & $2.38$/$-2.09$ & $0.09$ \\ 
&  & PBEsol & $16$ & $2.20$/$-1.58$ & $0.09$ \\ 
&  & QNA & $-27$ & $2.26$/$-1.82$ & $0.09$ \\ \hline
Exp. \cite{Romero2018} & FM &&& $2.67$ & $0.27$\\
& FM && $-34$ && \\
& FM && $-96$ && \\
 \hline \hline
\end{tabular}
\label{table:Fe3Pt_results}
\end{table}
As a third example we follow Ref. \onlinecite{Romero2018} and calculate the formation energies and magnetic moments of ferromagnetic and paramagnetic Fe$_3$Pt in L$1_2$ structure.
In the ferromagnetic state all moments point in the same direction and in order to simulate the paramagnetic state one of the Fe moments is inverted with respect to the other two Fe atoms in the unit cell.
We used Fe and Pt PAW-setups similar to Ref. \onlinecite{Romero2018}, where $3d74s1$ and $5d96s1$ are treated as valence electrons for Fe and Pt, respectively.
We also tried both the ``MixerSum'' and ``MixerDif'' density mixing methods that GPAW offers, because in some cases there is a difference between the magnetic states (and the ground-state energies) to which the two mixers converge.
Table \ref{table:Fe3Pt_results} shows our results calculated with four different XC functionals, which are LDA, PBE, PBEsol, and QNA.
Calculations were run using two different density mixers mentioned above and in each case the results in the table correspond to whichever mixer that yielded the lower ground-state energy.
Unlike Ref. \onlinecite{Romero2018}, all four XC functionals predict the ferromagnetic state (FM) to be more stable than the paramagnetic state (PARA), although for LDA the difference between FM and PARA formation energies is very small.
The magnitudes of LDA formation energy and magnetic moments are underestimated compared to available experimental values.
PBE and PBEsol predict FM formation energies that are between the two experimental values, but PBEsol magnetic moments are slightly closer to experiments than those of PBE.
QNA predicts a formation energy that slightly overestimates the available experimental data, but like PBEsol the QNA magnetic moments are in very good agreement with experiments.
Overall, we can say that our QNA implementation is viable also for magnetic alloys.

\section{Conclusions}
We have implemented the Becke fuzzy cells type space partitioning scheme in GPAW for the purposes of the flexible GGA-level QNA exchange-correlation functional and tested its functionality for a few test systems.
In general, space partitioning allows one to define atomic site specific quantities or to divide the system at hand into physically different regions, such as a bulk region and surface regions.
Since the bulk and surface regions could now be calculated with separate exchange-correlation functionals that are specifically designed to capture the important physics of those regions, space partitioning is one possible route to improved DFT accuracy.

\section{Acknowledgments}
MK aknowledges Academy of Finland Grant number 295602.
The computer resources of the Finnish IT Center for Science (CSC) and the
Finnish Grid and Cloud Infrastructure (FGCI) project (Finland), and the Swedish National Infrastructure for Computing (SNIC) at the High Performance Computing Center North (HPC2N) are acknowledged.

\appendix
\section{Equations}
By taking the partial derivative with respect to nuclear position of Eq.~(\ref{Excpaw}), the QNA XC force contribution within the PAW formlism is of the form
\begin{align} 
{\bf F}_{\rm XC}^a =& -\frac{ {\partial} E_{\rm XC}[\tilde n] }{{\partial} {\bf R}^a} \nonumber \\
=& 
-
\int {\rm d}{\bf r}
\left(
\frac{\delta
E_{\rm XC}[\tilde n({\bf r}), \tilde \mu({\bf r}), \tilde \beta({\bf r})] }{\delta \tilde n({\bf r})}
\frac{\partial \tilde n({\bf r})}{\partial {\bf R}^a}
+\frac{\delta
E_{\rm XC}[\tilde n({\bf r}), \tilde \mu({\bf r}), \tilde \beta({\bf r})]}{\delta \tilde \mu({\bf r})}\right.
\frac{\partial \tilde \mu({\bf r})}{\partial {\bf R}^a} \nonumber \\
&+ \left.\frac{\delta
E_{\rm XC}[\tilde n({\bf r}), \tilde \mu({\bf r}), \tilde \beta({\bf r})]}{\delta \tilde \beta({\bf r})}
\frac{\partial \tilde \beta({\bf r})}{\partial {\bf R}^a} \right) \nonumber \\
&-
\sum_a \sum_{ii'}
\int {\rm d}{\bf r}
\left(
\frac{\delta E^a_{\rm XC}[n^a({\bf r}), \mu^a({\bf r}), \beta^a({\bf r})]}{\delta n^a({\bf r})}
\frac{\partial n^a({\bf r})}{\partial D^a_{ii'}} +
\frac{\delta E^a_{\rm XC}[\tilde n^a({\bf r}), \mu^a({\bf r}), \beta^a({\bf r})]}{\delta \tilde n^a({\bf r})}
\frac{\partial n^a({\bf r})}{\partial D^a_{ii'}}
\right)
\frac{\partial D^a_{ii'}}{\partial {\bf R}^a} \nonumber \\
&-\sum_a \sum_{ii'} 
\int {\rm d}{\bf r}
\left(
\frac{\delta E^a_{\rm XC}[n^a({\bf r}), \mu^a({\bf r}), \beta^a({\bf r})]}{\delta \mu^a({\bf r})}
\frac{\partial \mu^a({\bf r})}{\partial D^a_{ii'}} +
\frac{\delta E^a_{\rm XC}[\tilde n^a({\bf r}), \mu^a({\bf r}), \beta^a({\bf r})]}{d \tilde \mu^a({\bf r})}
\frac{\partial \tilde \mu^a({\bf r})}{\partial D^a_{ii'}}
\right)
\frac{\partial D^a_{ii'}}{\partial {\bf R}^a}
\nonumber \\
&+\sum_a \sum_{ii'} 
\int {\rm d}{\bf r}
\left(
\frac{\delta E^a_{\rm XC}[n^a({\bf r}), \mu^a({\bf r}), \beta^a({\bf r})]}{d \beta^a({\bf r})}
\frac{\partial \beta^a({\bf r})}{\partial D^a_{ii'}} +
\frac{\delta E^a_{\rm XC}[\tilde n^a({\bf r}), \mu^a({\bf r}), \beta^a({\bf r})]}{d \tilde \beta^a({\bf r})}
\frac{\partial \tilde \beta^a({\bf r})}{\partial D^a_{ii'}}
\right)
\frac{\partial D^a_{ii'}}{\partial {\bf R}^a}.
\label{eq:dE_QNA_dR}
\end{align}

In Eqs. (\ref{eq:dEXC_dmu}) and (\ref{eq:dEXC_dbeta}) the partial derivatives
\[
\frac{\partial\left\{
\tilde n({\bf r})\varepsilon_{\rm XC}^{\rm PBE}[\tilde n({\bf r}), |\nabla \tilde n({\bf r})|^2, \tilde \mu({\bf r}), \tilde \beta({\bf r})]\right\}
}{\partial \tilde \mu({\bf r})},\quad
\frac{\partial\left\{
\tilde n({\bf r})\varepsilon_{\rm XC}^{\rm PBE}[\tilde n({\bf r}), |\nabla \tilde n({\bf r})|^2, \tilde \mu({\bf r}), \tilde \beta({\bf r})]\right\}
}{\partial \tilde \beta({\bf r})},
\]
when written out, become
\begin{align}
&\frac{\partial\left\{
\tilde n({\bf r})\varepsilon_{\rm XC}^{\rm PBE}[\tilde n({\bf r}), |\nabla \tilde n({\bf r})|^2, \tilde \mu({\bf r}), \tilde \beta({\bf r})]\right\}
}{\partial \tilde \mu({\bf r})} \nonumber \\
=& 
\frac{\partial \left[
\tilde n({\bf r})\varepsilon_{\rm X}^{\rm LDA} \left(
F_{\rm X}^{\rm PBE}[\tilde n({\bf r}), |\nabla \tilde n({\bf r})|^2, \tilde \mu({\bf r})] +
\left\{ \frac{\varepsilon_{\rm C}^{\rm LDA}}{\varepsilon_{\rm X}^{\rm LDA}} + \frac{H[\tilde n({\bf r}), |\nabla \tilde n({\bf r})|^2, \tilde \beta({\bf r})]}{\varepsilon_{\rm X}^{\rm LDA}}\right\} \right) \right]} {\partial \tilde \mu({\bf r})} \nonumber \\
=&\tilde n({\bf r})\varepsilon_{\rm X}^{\rm LDA} \frac{\partial F_{\rm X}^{\rm PBE}}{\partial \tilde \mu}
= \tilde n({\bf r})\varepsilon_{\rm X}^{\rm LDA} \left[ \frac{s^2}{(1+\mu s^2/\kappa)^2} \right],
\end{align}
\begin{align}
&\frac{\partial\left\{
\tilde n({\bf r})\varepsilon_{\rm XC}^{\rm PBE}[\tilde n({\bf r}), |\nabla \tilde n({\bf r})|^2, \tilde \mu({\bf r}), \tilde \beta({\bf r})]\right\}
}{\partial \tilde \beta({\bf r})} \nonumber \\
=&\frac{\partial \left[
\tilde n({\bf r})\varepsilon_{\rm X}^{\rm LDA} \left(
F_{\rm X}^{\rm PBE}[\tilde n({\bf r}), |\nabla \tilde n({\bf r})|^2, \tilde \mu({\bf r})] +
\left\{ \frac{\varepsilon_{\rm C}^{\rm LDA}}{\varepsilon_{\rm X}^{\rm LDA}} + \frac{H[\tilde n({\bf r}), |\nabla \tilde n({\bf r})|^2, \tilde \beta({\bf r})]}{\varepsilon_{\rm X}^{\rm LDA}}\right\} \right) \right]} {\partial \tilde \beta({\bf r})} \nonumber \\
=&\tilde n({\bf r}) \frac{\partial H}{\partial \beta} = \frac{Y t^2}{X \gamma} \left[ \frac{1+2At^2}{1+At^2+A^2t^4} - \frac{(1+At^2)(At^2+2A^2t^4)}{(1+At^2+A^2t^4)^2}  \right],
\end{align}
where we have used the notation of Ref. \onlinecite{Perdew1996a} as
\begin{align}
Y &= \frac{e^2}{a_0}\gamma\phi^3, \\
H &= Y\times \ln\left\{ 1 + \frac{\beta}{\gamma}t^2\left[ \frac{1+At^2}{1+At^2+A^2t^4} \right] \right\}, \\
X &= 1 + \frac{\beta}{\gamma}t^2\left[ \frac{1+At^2}{1+At^2+A^2t^4} \right], \\
A &= \frac{\beta}{\gamma} \left[ \exp{ \left\{ -\varepsilon_{\rm c}^{\rm LDA}/Y \right\} -1 } \right]^{-1}, \\
\frac{\delta A}{\delta \beta} &= \frac{A}{\beta}.
\end{align}
Above $e$ is the elementary charge, $a_0$ is the Bohr radius, $\gamma = (1 - \ln 2)/\pi^2$, and $\phi = [ (1 + \zeta)^{2/3} + (1 - \zeta)^{2/3}]/2$, where $\zeta = (n_\uparrow - n_\downarrow)/n$ is the relative spin polarization.

\section{Computational details}
The analytical force and stress tensor test used the planewave mode with an energy cutoff of \SI{600}{\electronvolt} and a $10\times10\times10$ grid of Monkhorst-Pack $k$-points \cite{Monkhorst1976}.
The version 0.9.20000 of PAW setups were used.
Fermi-Dirac smearing was used with a width of \SI{0.01}{\electronvolt}.

The ordered Cu-Au calculations used the planewave mode and an energy cutoff of \SI{550}{\electronvolt}.
Monkhorst-Pack scheme was used to generate the $k$-point grids whose sizes were $20\times20\times20$.
The version 0.9.20000 of PAW setups were used.
Fermi-Dirac smearing was used with a width of \SI{0.01}{\electronvolt}.

The 32-atom CuAu SQS calculations used the planewave mode and an energy cutoff of \SI{550}{\electronvolt}.
Forces were relaxed until the largest remaining force was smaller than $\SI{0.01}{\electronvolt}$, which ensured that the mixing energies were converged.
A $10\times10\times10$ grid of Monkhorst-Pack $k$-points was used.
The version 0.9.20000 of PAW setups were used.
Fermi-Dirac smearing was used with a width of \SI{0.01}{\electronvolt}.

The Fe$_3$Pt calculations used planewave basis and an energy cutoff of \SI{600}{\electronvolt}.
We used Fe and Pt PAW-setups similar to Ref. \onlinecite{Romero2018}, where $3d74s1$ and $5d96s1$ are treated as valence electrons for Fe and Pt, respectively.
We used $10\times10\times10$ and $15\times15\times15$ Monkhorst-Pack $k$-point grids for L$1_2$ (Fe$_3$Pt, bulk Pt) and B$_2$ (bulk Fe) structures, respectively.
Fermi-Dirac smearing was used with a width of \SI{0.1}{\electronvolt}.
MixerSum and MixerDif density mixers used the following settings: \{backend: pulay, beta: 0.02, nmaxold: 1, weight: 100\}.

%

\end{document}